\documentclass[journal=jacsat,manuscript=article]{achemso}

\usepackage{chemformula} 
\usepackage[T1]{fontenc} 

\usepackage[version=4]{mhchem}
\usepackage{amsfonts}
\usepackage{amsmath}

\usepackage{graphicx}

\usepackage{color}
\usepackage{graphicx}
\usepackage{lineno}
\usepackage{epstopdf}
\usepackage{amssymb}

\usepackage{longtable}

\usepackage{bm}
\usepackage{epstopdf}
\usepackage{graphicx}
\newcommand{\addfig}[3]{
	\begin{figure}[H]
		\centering
		\includegraphics[width=1\textwidth]{figures/#1}
		\caption{#2}
		\label{#3}
	\end{figure}
	}

\newcommand{\eq}[1]{(\ref{#1})}
\newcommand{\eqs}[2]{(\ref{#1}-\ref{#2})}
\newcommand{\reftab}[1]{Table \ref{#1}}
\newcommand{\reffig}[1]{Figure \ref{#1}}

\renewcommand{\vec}[1]{\boldsymbol{#1}}

\newcommand{\mat}[1]{\mathbf{#1}}

\newcommand{\Cdot}{\text{\tiny \textbullet}}

\newcommand{\vo}[0]{\vec{0}}

\newcommand{\vtheta}[0]{\vec{\theta}}

\newcommand{\iv}[0]{i^{(v)}}
\newcommand{\jv}[0]{j^{(v)}}
\renewcommand{\ln}[1]{\textbf{ln}\left(#1\right)}

\newcommand{\vlambda}[0]{\mathbf{\vec{\lambda}}}
\newcommand{\vepsilon}[0]{\mathbf{\vec{\epsilon}}}
\newcommand{\vpsi}[0]{\mathbf{\vec{\psi}}}

\newcommand{\vi}[0]{\mathbf{\vec{i}}}
\newcommand{\vl}[0]{\mathbf{\vec{l}}}

\newcommand{\mSigma}[0]{\mat{\Sigma}}
\newcommand{\mSigmaK}[0]{\mSigma^{(K)}}
\newcommand{\mPsi}[0]{\mat{\Psi}}

\newcommand{\mE}[0]{\mat{E}}
\newcommand{\mI}[0]{\mat{I}}
\newcommand{\mL}[0]{\mat{L}}
\newcommand{\mS}[0]{\mat{S}}
\newcommand{\mT}[0]{\mat{T}}
\newcommand{\mU}[0]{\mat{U}}
\newcommand{\mV}[0]{\mat{V}}
\newcommand{\mX}[0]{\mat{X}}
\newcommand{\mY}[0]{\mat{Y}}

\newcommand{\vx}[0]{\vec{x}}

\newcommand{\mW}[0]{\mat{W}}

\newcommand{\vy}[0]{\vec{y}}

\graphicspath{{./figures/}}

\author{Stefania Russo}
\email{stefania.russo@eawag.ch}
\affiliation[Eawag]
{Eawag, Swiss Federal Institute of Aquatic Science and Technology, 8600 D{\"u}bendorf, Switzerland}
\author{Guangyu Li}
\email{ligu@student.ethz.ch}
\affiliation[Eawag]
{Eawag, Swiss Federal Institute of Aquatic Science and Technology, 8600 D{\"u}bendorf, Switzerland}
\alsoaffiliation[ETHZ]
{ETH Z{\"u}rich, Institute of Environmental Engineering, 8093 Z{\"u}rich, Switzerland}
\author{Kris Villez}
\email{kris.villez@eawag.ch}
\affiliation[Eawag]
{Eawag, Swiss Federal Institute of Aquatic Science and Technology, 8600 D{\"u}bendorf, Switzerland}
\alsoaffiliation[ETHZ]
{ETH Z{\"u}rich, Institute of Environmental Engineering, 8093 Z{\"u}rich, Switzerland}

\title[Dimensionality selection with ignorance score]
  {Automatic dimensionality selection for principal component analysis models with the ignorance score}

\begin{document}

%

\begin{abstract}
Principal component analysis (PCA) is by far the most widespread tool for unsupervised learning with high-dimensional data sets. Its application is popularly studied for the purpose of exploratory data analysis and online process monitoring. Unfortunately, fine-tuning PCA models and particularly the number of components remains a challenging task. Today, this selection is often based on a combination of guiding principles, experience, and process understanding. Unlike the case of regression, where cross-validation of the prediction error is a widespread and trusted approach for model selection, there are no tools for PCA model selection which reach this level of acceptance. In this work, we address this challenge and evaluate the utility of the cross-validated ignorance score with both simulated and experimental data sets. Application of this method is based on the interpretation of PCA as a density model, as in probabilistic principal component analysis, and is shown to be a valuable tool to identify an optimal number of principal components.
\end{abstract}

\section{Introduction}

Principal component analysis (PCA) is one of the most popularly applied and studied methods within the context of data mining, machine learning, and process monitoring \cite{Joliffe2002,Yin2014}. 
This is partially explained by the existence and widespread implementation of efficient algorithms for data decomposition by PCA. Despite its widespread use, it remains difficult to provide universal guidelines for the optimization of the single hyper-parameter of PCA models, i.e. the number of principal components (PCs) that ought to be retained. One of the most important challenges is that PCA models can be hard to interpret. In addition, the use of PCA implies that retaining more variance means capturing more information, an assumption that is often hard to inspect carefully. The problem of selecting an optimal model hyper-parameter is shared with many unsupervised learning methods \cite{Smyth2000,Xiao2015,Wang2018}

The adequate selection of the number of PCs is crucial for both applications relying on data compression (e.g., exploratory analysis) as well as for predictive tasks (e.g., multivariate classification and regression). As a result, a wide variety of criteria for determining the number of PCs have been proposed \cite{Joliffe2002}. A first group of criteria are grounded in theory  \cite{Ferre1995,Minka2001,Bouveyron2011,Josse2012}. 
In general, such theoretical criteria assume that the linear PCA model with spherical Gaussian measurement noise and Gaussian distributions for the latent variables is adequate. Several authors have proposed cross-validation methods \cite{Valle1999,Bro2008}, inspired by the success of cross-validation procedures in regression and classification model identification \cite{kohavi1995study}. We continue along this line of research primarily because cross-validation as a method for model selection does not require that the assumed model structure is correct.


The available cross-validation methods for PCA model selection can be divided along the two cross-validation pattern that are in use. When using the first pattern, one builds a model with a subset of the data set (calibration set) and test the PCA model on one or more samples that were not used during calibration (validation set). This is known as row-wise cross-validation or RKF. Importantly, applying RKF means a complete validation data sample is used at once during model performance testing. In recent years, the RKF pattern has been criticized as a tool for PCA cross-validation because any cross-validated performance criterion, such as the sum-of-squares of prediction error, always exhibits a monotonic profile as function of the number of PCs \cite{Saccenti2015a}. This is a result of using the validation data to first compute the associated principal scores and then using these data again to compare with the reconstruction on the basis of the computed scores. This can be considered a form of data leakage \cite{Smialowski2009}. 
This is unlike the case of regression, where model selection can be based on identifying the model with a minimal cross-validated prediction error without use of the predicted values to obtain the model predictions. It is for this reason primarily that the element-wise cross-validation pattern (EKF) is attractive. In this case, one removes a number of elements from a validation sample and then uses the identified PCA model and the remaining measurements in the validation sample to impute these removed values. In this case, there is no data leakage and the profile of the cross-validated criterion, e.g. the reconstruction error, is likely to exhibit a minimum leading to a straightforward selection of a PCA model with an optimal number of PCs. One straightforward approach consists of imputing every element separately for every validation sample. One important inconvenience of EKF-based cross-validation is that the computational cost does not scale well with the number of variables in a data set. 


Recent evidence \citep{Camacho2012,Camacho2014,Saccenti2015,Saccenti2015a} suggests that the exact method to determine the optimal number of PCs may depend strongly on the intended use of the PCA models. For example, finding a minimal model for missing data imputation is best approached with cross-validation based on trimmed score imputation (TRI) combined with a corrected EKF pattern \cite{Camacho2014}. In this work, we refer to this method as PCA EKF cTRI. This reflects that the applied correction primarily relates to the imputation method. The EKF pattern is applied the same with both TRI and cTRI. When a low dimensionality is not a strict requirement and only the quality of the imputed data is of concern, trimmed score regression (TSR) is a suitable imputation method \cite{Camacho2014}. To date, the RKF pattern is still recommended for applications relying on data compression \cite{Camacho2014}, despite the fact that existing RKF-based cross-validated criteria for PCA models lead to data leakage and cannot not produce a minimum when plotted against the number of PCs \cite{Saccenti2015,Saccenti2015a}. Thus, PC dimensionality selection remains challenging for the applications relying on data compression by PCA, such as exploratory data analysis \cite{Villez2008a,Maere2012} and process and system monitoring \cite{Wise1990,Tong1995,Ramaker2004,Perelman2012,Villez2016,Zhang2017}. For such applications, the literature does not provide evidence in favor of a cross-validation method which is simultaneously accurate, automated, efficient, and intuitive. The search for such a method is therefore the focus of this work.

In the following sections, we describe the simulated and experimental data set used in this work. Then, we discuss the PCA EKF cTRI cross-validation method and our newly proposed method based on the application of the ignorance score \cite{Gneiting2007} to probabilistic principal component analysis (PPCA) \cite{Tipping1999a}. We proceed with a classical structure including results, discussion, and conclusions.

\section{Materials and Methods}

We first describe the simulated and experimental data sets used for this study. After, the applied methods for data analysis are described in detail. All mathematical symbols are listed in \reftab{tab:symbol}.

\begin{center}
\small
\begin{longtable}{p{3.5cm}p{6cm}p{3cm}}
\caption{List of sensor characterization tests} \label{tab:symbol} \\
\hline
      \textbf{Symbol} & \textbf{Description} & \textbf{Dimensions}\\
\hline
\endfirsthead
\multicolumn{3}{c}%
{\tablename\ \thetable\ -- \textit{Continued from previous page}} \\
\hline
      \textbf{Symbol} & \textbf{Description} & \textbf{Dimensions}\\
\hline
\endhead
\hline \multicolumn{3}{r}{\textit{Continued on next page}} \\
\endfoot
\hline
\endlastfoot
		$\vepsilon$ 
	&	Vector of measurement errors
	&	$J \times 1$
\\ 		$\vtheta$
	&	Distribution parameters
	&	$ $
\\ 		$\vlambda$
	&	Vector of eigenvalues
	&	$K \times 1$
\\ 		$\mSigmaK$
	&	Covariance matrix estimate
	&	$J \times J$
\\		$\sigma_{\epsilon}$
	&	Estimated noise variance
	&	$1 \times 1$
\\		$\phi$
	&	Expected variance of imputed value
	&	$1 \times 1$
\\		$\mPsi$
	&	Diagonal matrix with latent variable variances on the diagonal
	&	$K \times K$
\\		$\vpsi$	
	&	Vector of latent variable variances
	&	$K \times 1$
\\		$\mE$
	& 	Imputation error
	&	$I \times J$
\\		$e$ 
	&	Noise level
	&	$1 \times 1$
\\		$\mI_J$
	&	Identity matrix
	&	$J \times J$
\\		$I$ $\left(I^{(c)}, I^{(v)}\right)$
	&	Number of samples / rows (for calibration, validation)
	&	$1 \times 1$
\\		$\vi^{(c)}$, $\vi^{(v)}$		
	&	Row numbers of calibration and validation samples
	&	$I^{(c)} / I^{(v)} \times 1$
\\		$i$ $\left( \iv \right)$
	&	Sample / row index  (for validation)
	&	$1 \times 1$
\\		$J$ 
	&	Number of variables / columns
	&	$1 \times 1$
\\		$j$ $\left( \jv \right)$					
	&	Variable / column index (for validation)
	&	$1 \times 1$
\\		$K$ $\left( K^{*}\right)$
	&	Number of principal components (maximum)
	&	$1 \times 1$
\\		$k$
	&	Index of principal component
	&	$1 \times 1$
\\		$\mL$
	&	Matrix of element-wise ignorance scores
	&	$I \times J$
\\		$\vl$
	&	Vector of whole-sample ignorance scores
	&	$I \times 1$
\\		$r$
	& 	Repetition number
	&	$1 \times 1$
\\		$\mS $
	&	Diagonal matrix of singular values
	&	$K^{*} \times K^{*}$
\\		$s$
	& 	Data set type 
	&	$1 \times 1$
\\		$\mT$ $\left( \mT^{(c)}, \mT^{(v)}\right) $
	&	Principal scores (of the calibration/validation data)
	&	$I / I^{(c)} / I^{(v)}  \times K^{*}$ 
\\		$\mT^{(c,aug)}$, $\mT^{(v,aug)}$
	&	Principal scores of the augmented calibration/validation data
	&	$I^{(c)} /I^{(v)}  \times K$ 
\\ 		$\mU$ $\left(\mU^{(c)}\right)$
	&	Standardized principal scores (of the calibration data)
	&	$I \times J$ $\left( I^{(c)} \times 1 \right)$
\\		$\mV$ ($\mV^{(aug)}$)
	&	Matrix of loading vectors (of the augmented data)
	&	$J \times K^{*}$ ($J \times K$)
\\		$\mW$
	&	Matrix of data-generating loading vectors
	&	$J \times K$
\\		$\mX$
	&  	Matrix of latent variables
	&	$I \times K$
\\		$\vx$
	&	Vector of latent variables
	&	$K \times 1$
\\		$\mY$ $\left(\mY^{(c)}, \mY^{(v)} \right) $
	&	Data matrix (for calibration, for validation)
	&	$I / I^{(c)} / I^{(v)} \times J$
\\		$\hat{\mY}^{(c)}$ , $\hat{\mY}^{(v)} $ 
	&	Approximation of calibration/validation data
	&	$I^{(c)} / I^{(v)} \times J$
\\		$\mY^{(c,aug)}$ $\left(\hat{\mY}^{(c,aug)}\right) $
	&	Augmented calibration data (approximation)
	&	$I^{(c)} \times (J+K) $
\\		$\vy$
	&	Vector of measured variables
	&	$J \times 1$
\\		$y$
	&	Scalar measurement
	&	$1 \times 1$
\end{longtable}
\end{center}

\subsection{Data sets}

\subsubsection{Simulation data sets}

Simulated data sets are constructed with a known number of latent variables. 
These simulation data sets are devised for the specific purpose of benchmarking of automated model dimensionality selection methods \cite{Camacho2014}. Each data set is indexed as $s.e$, where $s$ indicates the data set types ( $s = 1\, \ldots \, 4$ )  and $e$ corresponds to the different relative measurement noise variances ( $e = 1\, \ldots\, 6$ ). The first 5 noise variance correspond to those used before ($5, 10, 15, 20, 25$\%) \cite{Camacho2014} whereas the last one equals $50$\%.  This results in a total number of 24 data sets. Each data set consists of a $I \times J$-dimensional matrix with $I$ always equal to 1024 and $J$ equal to 10, 10, 27, and 50 for $s$ ranging from 1 to 4. The known numbers of latent variables are 8, 8, 12, and 15.  The procedure to simulate each of these data sets is detailed in the {\em Supplementary Information } and is repeated 100 times ( $r=1 \, \ldots \, 100$ ). An individual repetition of a particular data set is indicated as $s.e.r$. For example, the 73rd instance of the data set of type 4 with noise level 5 is indicated with $4.5.73$.


\subsubsection{Laboratory data set}

To test the quality of the studied methods under real-world conditions, experimental absorbance spectra are collected in such a way that a PCA model with two PCs is expected to describe the obtained data well. To this end, stock solutions of nitrite (\ce{NO_{2}^-}) and nitrate (\ce{NO_{3}^-}) are prepared first. Each of these two stock media were prepared as a single batch with target concentrations of 5 g atomic nitrogen per liter (5 \ce{NO_{2}^{-}}-N/L, 5 g\ce{NO_{3}^{-}}-N/L). 

With the prepared stock solutions, diluted media were obtained adding 600 mL of nano-filtered water to a glass cylinder first. Then, well-measured amounts of the two stock solutions are added in steps of 2 mL with a minimum of 1 and a maximum of 16 steps, leading to a square two-dimensional grid of added volumes of the two stock solutions as shown in \reffig{fig:blocks}. As a result, the expected concentrations of both species range from 16.6 mg N/L (1 step addition with both stock solutions) to 252.4 mg N/L (e.g., 1 step with \ce{NO_{2}^{-}} and 16 steps with \ce{NO_{3}^{-}} stock solution). The order in which the samples were prepared was randomized partially to avoid temporal auto-correlation within the collected data sets. More details regarding the experimental procedure used to prepare the solutions are found in the {\em Supplementary Information}.

Immediately after preparation of each diluted solution, five replicate absorbance spectra were collected by submerging an on-line ultraviolet-visible light absorbance spectrophotometer (spectro::lyser$^{TM}$, S::CAN Messtechnik, Vienna, Austria) into the diluted solution. The applied spectrophotometer has a light path length of 2 mm and produces measurements which are composed of 215 absorbance values taken at wavelengths between 200 nm and 735 nm with steps of 2.5 nm. PCA models are studied for a variety of variable selections for reasons explained below. We refer to the resulting modified data sets as laboratory data set 0 (no variable selection, 215 wavelengths), laboratory data set 1 (wavelengths 285-735 nm, 181 wavelengths), and laboratory data set 2 (wavelengths 285-385 nm, 41 wavelengths).

The design of the experiment includes two factors, the nitrite and nitrate concentration. According to the Beer-Lambert law, the absorbance measurements are expected to depend linearly on these concentrations. Thus, a model with two components ought to deliver a good representation of the collected data. In addition, the number of factors in the experimental design can be used as a gold standard for evaluation of automatic methods for latent variable model selection. 

\addfig{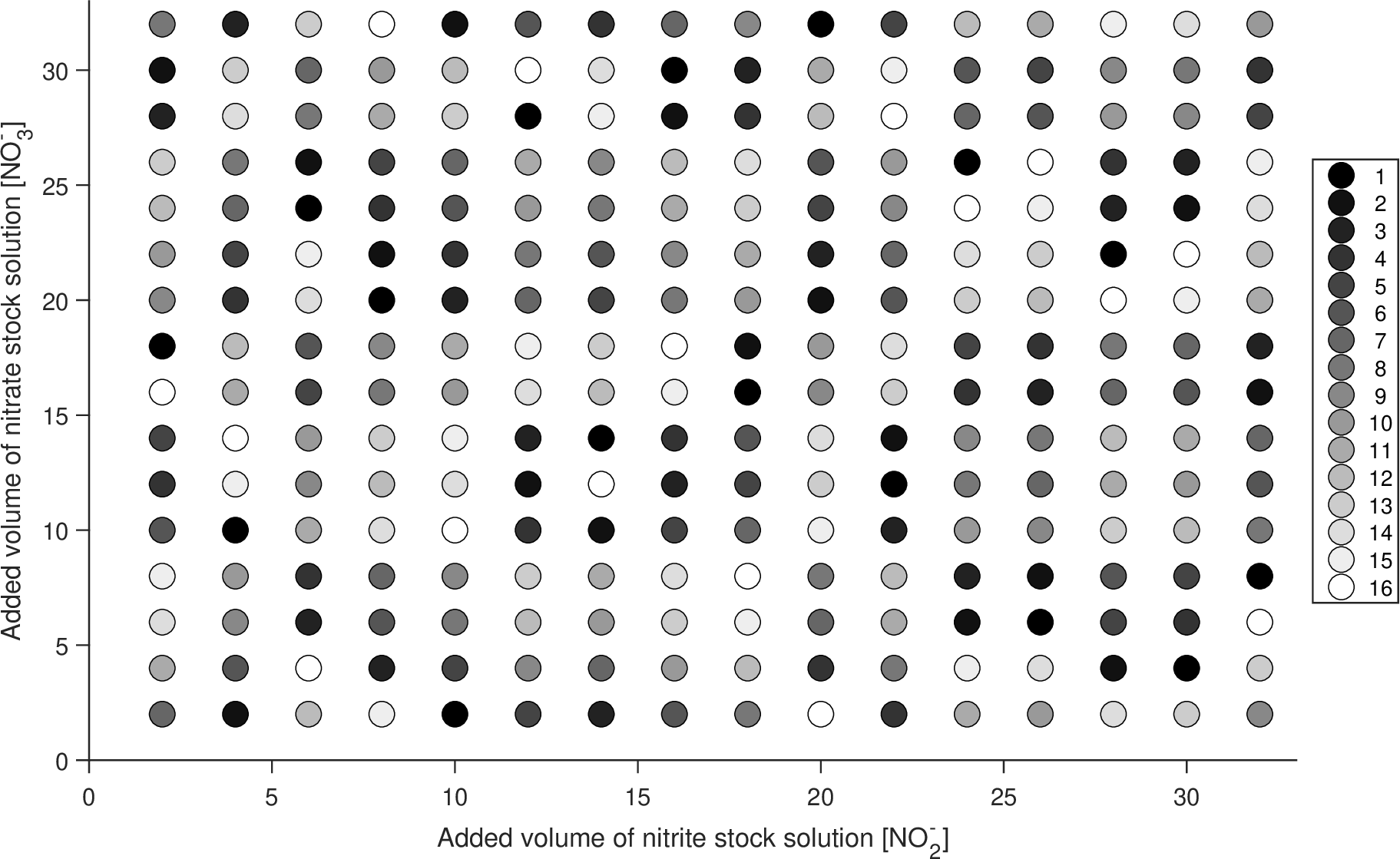}{{\bf Laboratory data set -- Experimental design and block assignment.} Five spectra are collected for each combination of added volume of the nitrite and nitrate stock solutions. The obtained spectra are divided in 16 blocks by means of a randomized Latin square for the purpose of cross-validation. Each block is indicated with a unique shade.}{fig:blocks}

\subsection{Methods}

We now explain how PCA and PPCA models are calibrated and validated in this study. The decomposition of the calibration data is a common step for both models and is explained first.

\subsubsection{Centering and scaling}

No centering or scaling is applied for the simulation data sets, as in previous studies \cite{Camacho2014}. In contrast, the laboratory data are mean centered before analysis. To this end, the column-wise mean is computed on the basis of the calibration data and subtracted from both calibration and validation data (defined below). In the present study, we apply no scaling at any time.

\subsubsection{Singular value decomposition}

We write the matrix containing all data (after centering, if applied) as $\mY$ with $I$ rows corresponding to data samples ( $i=1 \, \ldots \, I$ ) and $J$ columns corresponding to measured variables ( $j=1 \, \ldots \, J$ ). The calibration data is given as obtained by selecting the $I^{(c)}$ ( $I^{(c)} < I$ )  rows from this matrix. The calibration data is obtained by selecting $I^{(c)}$ rows in the data matrix ($I^{(c)}<I$ with $\vi^{(c)}$ the $I^{(c)}$-dimensional vector containing the row numbers for these samples: $\mY^{(c)}=\mY_{\vi^{(c)},\Cdot}$. In the present work, we assume that there are more samples and variables in the calibration data set ($I^{(c)}>J$). Consequently, we define the maximal number of principal components as $K^{*} := \min(I^{(c)},J)=J$. 

The calibration data matrix is decomposed with singular value decomposition (SVD) so that:

\begin{align}
		\mY^{(c)}  &= \mU^{(c)} \, \mS \, \mV^{T} := \mT^{(c)} \, \mV^{T} 
\end{align}

with $\mT^{(c)}$ the $I^{(c)} \times K^{*}$-dimensional matrix of principal scores, $\mU^{(c)}$ the $I^{(c)} \times K^{*}$-dimensional matrix of standardized principal scores, $\mS$ the $K^{*} \times K^{*}$-dimensional diagonal matrix containing all non-zero singular values, ordered from largest to smallest, and $\mV$ the $J \times K^{*}$-dimensional matrix containing the loading vectors as columns.

\subsubsection{Principal component analysis}

\paragraph{Calibration}

Given the SVD result, principal component analysis (PCA) proceeds by selecting a number $K$ ( $K \leq K^{*}$ )  of components by choosing the first $K$ columns in $\mU^{(c)}$ and $\mV$ and selecting the first $K$ rows and columns of $\mS$. This leads to the following least-squares optimal approximation of the calibration data \cite{Schuermans2005}:

\begin{align}
	\mY^{(c)} & \approx  \hat{\mY}^{(c)} := \mU_{\Cdot,1:K}^{(c)} \cdot \mS_{1:K,1:K} \cdot {\mV_{\Cdot,1:K}}^T = \mT_{\Cdot,1:K}^{(c)} \cdot {\left( \mV_{\Cdot,1:K} \right)}^T \, .
\end{align}

Under the assumption of spherical multivariate normal measurement noise, this approximation corresponds to maximum likelihood principal component analysis \cite{Hoefsloot2006,Narasimhan2008,Wentzell1997,Wentzell1999}.  

The PCA model is validated on the basis of the corrected trimmed score imputation (cTRI) procedure \cite{Camacho2014}. To this end, an augmented PCA model is built by {\em (a)} constructing the following $I^{(c)} \times (J+K)$-dimensional augmented data matrix:

\begin{align}
	\mY^{(c,aug)} & = \left[ \begin{array}{cc} \mY^{(c)} & \mT^{(c)}_{\Cdot,1:K} \end{array} \right]
\end{align}

and {\em (b)} constructing an augmented PCA model with $K$ components by applying mean centering (if applied originally), SVD, and PCA calibration as outlined above. This then leads to the following PCA model for the augmented data:

\begin{align}
	\mY^{(c,aug)} & \approx \hat{\mY}^{(c,aug)} := \mT^{(c,aug)} \cdot {\left( \mV^{(aug)}\right)}^T
\end{align}

with $\mT^{(c,aug)}$ the $I^{(c)} \times K$-dimensional matrix of scores and $\mV^{(aug)}$ the $(J+K)\times K$-dimensional matrix of loading vectors.

\paragraph{Column-wise validation}

The $I^{(v)} \times J$-dimensional validation data matrix $\mY^{(v)}$ is obtained by selecting the $I^{(v)}$ rows in $\mY$ which are not in the calibration data matrix, $\vi^{(v)}$ $(I^{(c)}+I^{(v)}=I)$:

\begin{align} 
	\mY^{(v)} := \mY_{\vi^{(v)},\Cdot}
\end{align}

Validation then proceeds by treating a single column, $\jv$, in the validation data matrix as missing and testing the ability of the augmented PCA model to reconstruct this variable with the trimmed score imputation method. Practically, one first computes the scores of the original validation data matrix without missing data:

\begin{align}
	\mT^{(v)} & := \mY^{(v)}  \cdot {\mV_{\Cdot,1:K}}
\end{align}

Then, one modifies the $j$th column from the validation data matrix by setting each element in this column equal to zero:

\begin{align}
	\mY^{(v)}_{\Cdot,\jv} & \leftarrow 0
\end{align}

The resulting trimmed scores obtained by the augmented PCA model then are:

\begin{align}
	\mT^{(v,aug)} & := \left[ \begin{array}{cc} \mY^{(v)} & \mT^{(v)}  \end{array} \right] \cdot {\mV^{(aug)}}
\end{align}

Finally, the imputed values for the $\jv$th variable are:

\begin{align}
	\hat{\mY}^{(v)}_{\Cdot,\jv} & := 	\mT^{(v,aug)} \cdot {\left( \mV_{\jv,\Cdot}^{(aug)} \right)}^T
\end{align}

The quality of imputation is evaluated by means of the reconstruction error:

\begin{align}
	{\mE}_{\vi^{(v)},\jv} & := 	\hat{\mY}^{(v)}_{\Cdot,\jv}  - \mY^{(v)}_{\Cdot,\jv} 
\end{align}

\subsubsection{Probabilistic principal component analysis}

\paragraph{Calibration}

PPCA was introduced with the goal of formulating a probabilistic version of PCA \cite{Tipping1999a}. In the case of PPCA, the model is expressed as the maximum likelihood estimate (MLE) of the following generative latent variable model: 

\begin{align}
	\label{generation1}
	\vx_k & \sim \mathcal{N}\left(\vo, \vpsi_k \right)
	\\
	\label{generation2}
	\vepsilon & \sim \mathcal{N}\left(\vo, \mI_J \right)
	\\
	\label{generation3}
	\vy & := \mW  \, \vx + \vepsilon
\end{align}

with $\vx$ the $K$-dimensional vector of latent variables, $\vx_k$, each one of them drawn from a univariate normal distribution ( $k = 1 \, \ldots \, K$ ), $\vepsilon$ a $J$-dimensional vector of measurement errors drawn from a spherical multivariate normal density, $\mI_J$ an identity matrix of appropriate dimensions, $\mW$ a $J \times K$-dimensional matrix with full column rank, and $\vy$ the $J$-dimensional vector of recorded measurements. During model calibration one has access to $I$ vectors $\vy$ which are organized as row vectors in a $I^{(c)} \times J$-dimensional data matrix.

As in PCA, model identification proceeds by choosing a number $K$ for the number of principal components. Then, starting with the SVD result, the density model for the measured data is formulated as the following multivariate normal distribution:

\begin{align}
	\label{generation}
	\vy & \sim \mathcal{N}\left(\vo, \mSigmaK \right)
\end{align}

where $\mSigmaK$ is the MLE of the variance-covariance matrix given $K$ components. To compute this matrix, one first obtains $\vlambda$,  the vector of the first $K$ eigenvalues of the empirical variance-covariance matrix, as:

\begin{align}
	\vlambda_k & := \frac{{\left(\mS_{k,k}\right)}^2}{I^{(c)}} \, & k&=1\, \ldots \, K
\end{align}

Then the MLE of $\mSigmaK$ is available via:

\begin{align}
	\mSigmaK & := \mV_{\Cdot,1:K} \, \mPsi \, {\left(\mV_{\Cdot,1:K} \right)}^T + \sigma_{\epsilon} \, \mI_{J}
	\\
	\sigma_{\epsilon} & := \frac{1}{K^*-K} \sum_{k=K+1}^{K^{*}} {\left(\mS_{k,k}\right)}^2
	\\
	\vpsi_k & := \vlambda_k - \sigma_{\epsilon}, \, k=1\, \ldots \, K
	\\
	\mPsi & := \textbf{diag}{\left( \vpsi \right)}
\end{align}

where $\mPsi$ is a diagonal matrix with $\vpsi$, the vector of $K$ deflated variance estimates for the K selected components, on its diagonal, and $\sigma_{\epsilon}$ the estimated noise variance. $\mV_{\Cdot,1:K}$ functions as the MLE of $\mW$.

One can compute the scaled scores for the $k$th principal component, $\mX_{\Cdot,k}$, with the variance equal to $\vpsi_k$ as:

\begin{align}
	\label{deflatescores}
	\mX_{\Cdot,k} & := \mU_{\Cdot,k} \, \vpsi_k = \mT_{\Cdot,k} \, \frac{\vpsi_k}{\vlambda_k}
\end{align}

The row vectors in $\mX$ are the MLE of $\vx$ in \eq{generation1}. These vectors can be used in \eq{generation3} to simulate new data samples with the identified score values yet with newly sampled noise vectors. This is used below to evaluate whether the assumption of spherical measurement noise is valid for the laboratory data set.

A key observation is that PPCA explicitly accounts for the fact that the principal scores computed with PCA are subject to noise. Under the assumptions of spherical measurement noise and given a choice for $K$, the deflated eigenvalues reflect the magnitude of variation in the principal components that is not interpreted as noise. This also means that the fraction of the variance associated with a particular principal component that is interpreted as meaningful, i.e. non-noisy, {\em (a)} will be lower than the fraction of explained variance obtained with PCA ( $\vpsi_{k} < \vlambda_k$, $k=1, \ldots \, K$, $K < K^{*}-1$ ) and {\em (b)} will increase with increasing values of $K$ ( $K_1 < K_2: {\left. \vpsi_{k} \right|}_{K=K_1} \leq {\left. \vpsi_{k} \right|}_{K=K_2}, k=1, \ldots \, K_2$ ). Note also that $K=J-1$ and $K=J$ deliver the same estimate for $\mSigmaK$, the only distinction being the interpretation of the $J$th loading vector as a noisy direction ($K=J-1$) or an informative direction ($K=J$).

\paragraph{Column-wise validation}

To validate the PPCA model, we make use of the ignorance score \cite{Gneiting2007}. This criterion has been used in hydrological modeling for ensemble model calibration \cite{Boucher2009} 
and has been proposed in data mining to identify the number of clusters in density-based cluster models \cite{Smyth2000}. 
The ignorance score is defined as the negative logarithm of the likelihood of a scalar measurement, $y$, given a likelihood function $L(\Cdot) $:

\begin{align}
	\mathcal{I}(y) &= -log\, L(y,\vtheta)
\end{align}

with $\vtheta$ the vector of parameters. The ignorance score has the advantage of being a negatively oriented score \cite{Gneiting2007}, so that its minimum value indicates the model which maximizes the predicted likelihood of a measurement.

To apply this score for validation of a PPCA model, we proceed as follows. As before, we treat a single column, $\jv$, in the validation data matrix as missing. The PPCA model is used to compute the density of the $\jv$ variable conditional to the other measurements. This density is a univariate normal distribution. The imputed values, i.e. the conditional means for $\mY^{(v)}_{\Cdot,\jv}$ are:

\begin{align}
	\hat{\mY}^{(v)}_{\Cdot,\jv} & := 	\mY^{(v)}_{\Cdot,-\jv} \, {\left(\mSigmaK_{-\jv,-\jv}\right)}^{-1} \cdot  (\mSigmaK_{-\jv,\jv})
\end{align}

with $-\jv$ indicating the inclusion of all $J$ variables except $\jv$. One profits from the full-rank nature of $\mSigmaK_{-\jv,-\jv}$ in the above. The variance of the conditional density is:

\begin{align}
	\phi & :=\mSigmaK_{\jv,\jv} -  (\mSigmaK_{j^{(v),-\jv}}) \, {\left(\mSigmaK_{-\jv,-\jv}\right)}^{-1} \cdot  (\mSigmaK_{-\jv,\jv})
\end{align}

Now the univariate normal likelihood is applied to evaluate the ignorance score for each of the missing data points indexed by $\iv$ ( $\iv \in \vi^{(v)}$ ):

\begin{align}
	\mL_{\iv,\jv} &:=\mathcal{I}(\mY_{\iv,\jv}) = -\ln{L \left(\mY^{(v)}_{\iv,\jv},	\hat{\mY}^{(v)}_{\iv,\jv}, \phi \right) }  \\
	& = \frac{1}{2} \, \left( \ln{2 \pi } +  \ln{\left(\phi \right) }
	+  \frac{\left( \hat{\mY}^{(v)}_{\iv,\jv} -   \mY^{(v)}_{\iv,\jv} \right)^2 }{\phi} \right)
\end{align}

where $\mL$ is the $I \times J$ matrix of element-wise ignorance scores. 

\paragraph{Whole-sample validation}

Extending the ignorance score to evaluate the quality of multivariate normal density prediction is fairly straightforward. In this case, the ignorance score simply follows from evaluating the density associated with the PPCA model for a validation sample \cite{Gneiting2007}. The ignorance score for a single validation row, $\iv$, in the data matrix then follows as:

\begin{align}
	\vl_{\iv} := 	\mathcal{I}\left(\mY_{\iv,\Cdot}\right) & := -\ln{L \left(\mY_{\iv,\Cdot},	\vo, \mSigmaK \right)}   \\
	& = \frac{1}{2 \, J} \, \left( J \, \ln{ 2 \pi }+ \ln{\left| \Sigma^{(K)} \right|  } + \mY_{i^{(v)},\Cdot}  \, {\left(\mSigmaK \right)}^{-1} \, {\left( \mY_{i^{(v)},\Cdot} \right)}^T \right)
\end{align}

with $\vl$ the $I$-dimensional vector of whole-sample ignorance scores. In this study, we divide by $J$ on the right hand side to produce values in the same scale as the element-wise ignorance scores. Note that the determinant $\left| \Sigma^{(K)} \right|$ is the same no matter which value for $K$ is chosen. As a result, the whole-sample ignorance score is equal to the Mahalanobis distance with the estimated variance-covariance matrix $\Sigma^{(K)}$, apart from a constant. This is discussed further in the Discussion section.

\subsection{Cross-validation and model selection}

\subsubsection{Row-wise cross-validation} 

Row-wise cross-validation (RKF) is only applied for the PPCA model and using the whole-sample validation described above. All results are obtained by splitting the data into 16 blocks along the row dimension, each containing the same number of samples. Each data block is used once as a validation data matrix, while using the remaining ones for calibration (16-fold cross-validation). 

In the case of the simulated data sets, the block assignment is randomized completely. In contrast, block assignment for the laboratory data sets proceeds by means of a randomized Latin square. The collected spectra are first organized according to the 16 $\times$ 16 grid with the two axes corresponding to the added volumes of the stock solutions (\ce{NO_{2}^-}, \ce{NO_{3}^-}). Each position in this grid corresponds to 5 repetitions of the same absorbance measurements. Each grid position is assigned a block index (1 to 16) in a random fashion while making sure that each block index appears exactly once in every row and column. The result of this is shown in \reffig{fig:blocks}. Each of the 16 data blocks now correspond to all spectra with the same block index (16 grid positions $\times$ 5 repetitions = 80 spectra per block). 

The cross-validated ignorance score is then obtained as the mean ignorance score:

\begin{align}
	\frac{1}{I} \sum_{i=1}^I \vl_{i}
\end{align}

In what follows we refer to the method for model selection which combines PPCA model calibration and validation with an RKF pattern and the ignorance score as PPCA RKF IGN.

\subsubsection{Element-wise cross-validation} 

Element-wise cross-validation (EKF) is applied for both the PCA and PPCA models, each time using the corresponding column-wise validation procedure describe above. In the row/sample direction, the block-wise approach as described above is used. In the column/variable direction, we apply a leave-one-out approach as dictated by the validation procedures described above. For the PCA model, this leads to a cross-validated mean-squared-error as the model selection criterion:

\begin{align}
	\frac{1}{I \, J} \sum_{i=1}^I	\sum_{j=1}^J	 {\mE}_{i,j}^2
\end{align}

Similarly, a cross-validated ignorance score is obtained for PPCA model selection:

\begin{align}
	\frac{1}{I \, J} \sum_{i=1}^I	\sum_{j=1}^J \mL_{i,j} 
\end{align}

In what follows we refer to the method for model selection which combines PPCA model calibration and validation with an EKF pattern and the ignorance score as PPCA EKF IGN. Similarly, the method based on PCA, EKF, and cTRI is described as PCA EKF cTRI.

\subsubsection{Model selection}

In all cases, we select the PCA or PPCA model which delivers the minimal value for the cross-validated model performance criterion.

\subsubsection{Benchmarking with simulated data set}

The efficiency and accuracy of the studied methods for model selection are evaluated with the simulated data sets. To quantify the performance of the model selection procedures, the following criteria are computed:
\begin{enumerate}
    \item The average run time, measured in seconds, for a single execution of the studied cross-validation procedure. To this end, all methods are executed on a single machine.
    \item The fraction of the total number of instances of a data set for which the identified number of PCs matches the ground truth value exactly. This is reported as a percentage.
\end{enumerate}

In addition, we inspect the distribution of the identified number of principal components as a function of the data set type and noise level.

\section{Results }

The results with simulated data are discussed first. After that, results obtained with the laboratory data set are discussed.

\subsection{Simulation data sets}

\subsubsection{Exemplary cross-validation results}

\reffig{fig:CVexample} shows the cross-validated criteria obtained with data set 4.6.73 (type 4, noise level 5, repetition 73). Each of these criteria exhibits a unimodal profile (i.e., single minimum) which makes the automatic selection of a number of PCs ($K$) particularly easy. In all cases, the selected number matches the simulated number of PCs exactly.

\addfig{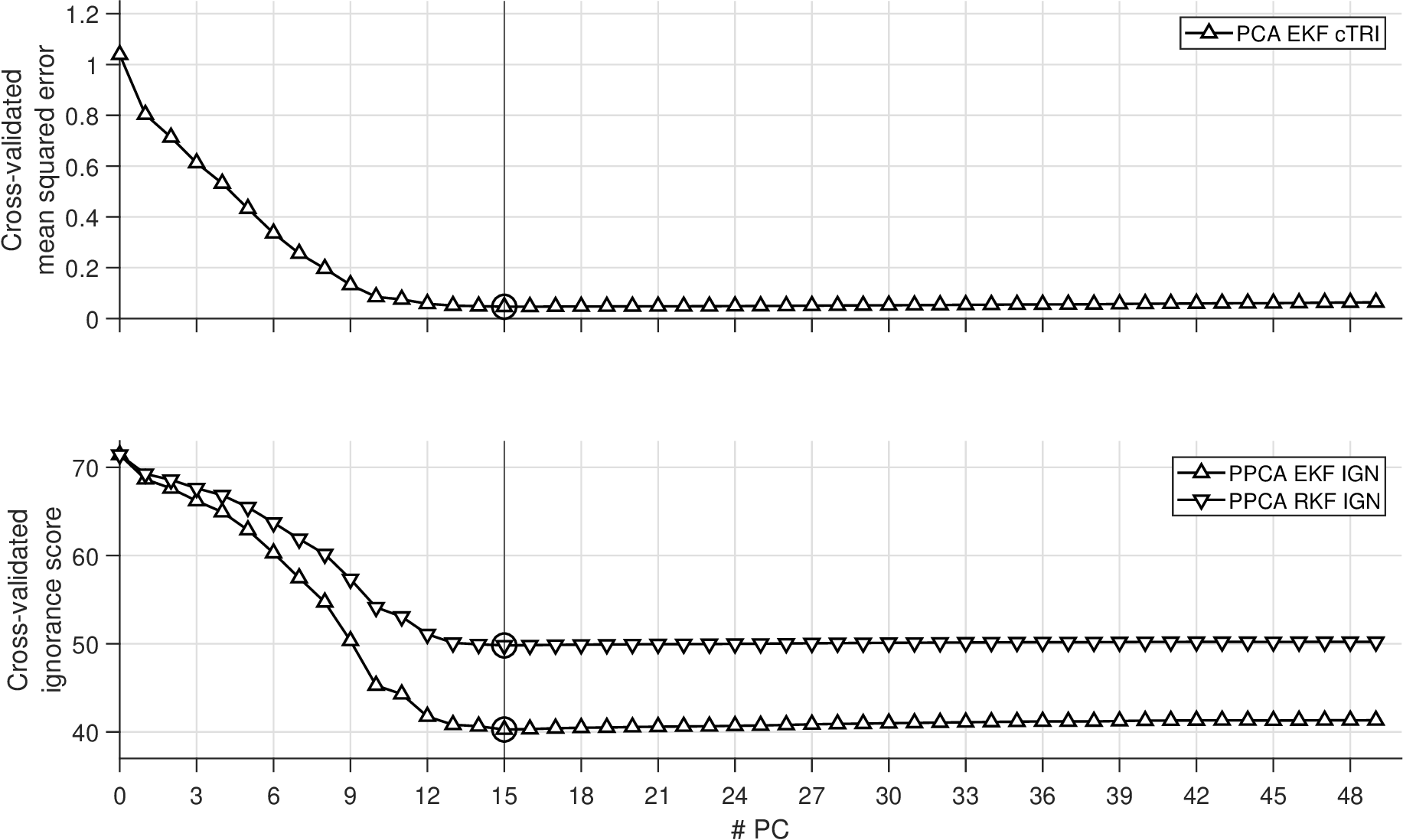}{{\bf Simulated data set 4.5.73 -- Typical cross-validation result.} Cross-validated criteria as a function of the number of retained principal components. In this case, all cross-validation methods pick the same number of principal components (circles) which equals the ground-truth value (vertical line).}{fig:CVexample}

\subsubsection{Benchmarking with simulated data sets}

Data has been simulated according to the 24 different data sets (1.1 until 4.6), each time generating 100 instances of the data set. \reffig{fig:CVaccuracy} shows the fraction of the instances where the identified number of PCs matches the simulated ground truth.
Not surprisingly, the lowest accuracy is found for all methods whenever the data contains the highest noise level. The PCA model with EKF-cTRI cross-validation leads to a 0\% accuracy for this highest noise level for all data set types. For noise levels 1 to 4, the PCA EKF cTRI method delivers the highest and almost perfect accuracy. At the highest noise level, this method is outperformed by the PPCA RKF IGN method delivering a slightly better performance than PPCA EKF IGN. At noise level 5, PCA EKF cTRI is best for data types 1 and 2 yet worst for data set types 3 and 4. The PPCA-based methods fare well for data set types 3 and 4 with an accuracy between 95\% and 100\% regardless of the applied noise level.

\addfig{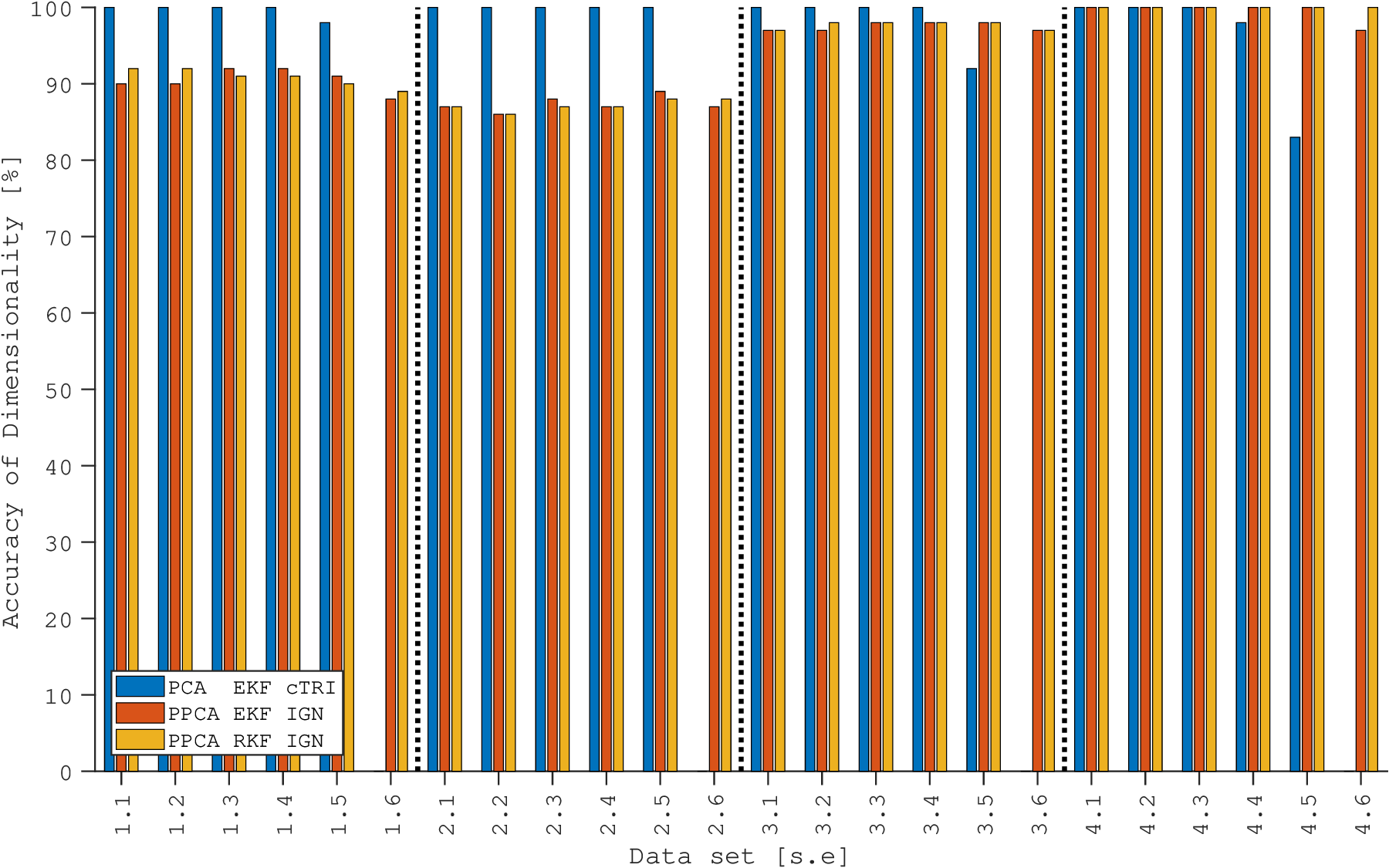}{{\bf Fraction of data set instances where the identified number of PCs matches the simulated ground truth.} The fraction of correctly identified number of principal components is shown for every data set (1.1 until 4.6) and every considered combination of model and cross-validation method. All methods achieve an accuracy of 80\% in all cases except with the PCA EKF cTRI method at the highest noise level.}{fig:CVaccuracy}

\begin{figure}[H]
		\centering
		\includegraphics[width=.73\textwidth, angle =+90]{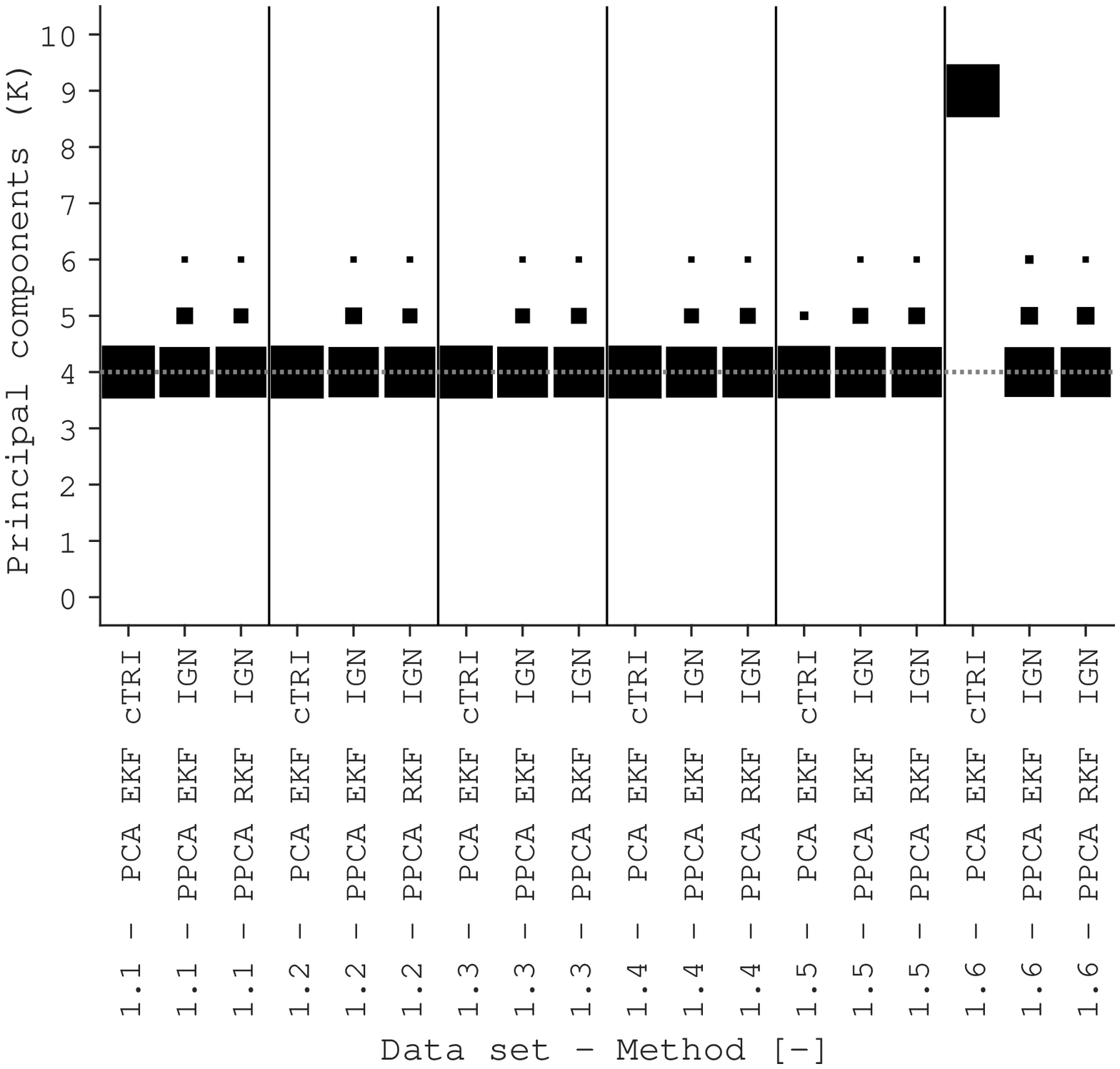}
		\caption{
		{\bf Distribution of identified numbers of PCs for the data sets 1.1 to 1.6.} For each number of principal components ( $K$, bottom to top), and for every noise level and every method (left to right), a black box is shown with a surface proportional to the number of data instances for which $K$ component are selected. The horizontal line indicates the simulated value of $K$ (5). With the PPCA-based methods, the identified number of PCs is never too low and at most 2 PC too high. With the PCA model, the results are perfect except for data set 1.5 and 1.6. With data set 1.6, PCA-EKF-cTRI the number of retained PCs always equals the maximum value that could be obtained (9).
		}
		\label{fig:CVdist1}
	\end{figure}

\begin{figure}[H]
		\centering
		\includegraphics[width=.73\textwidth, angle =+90]{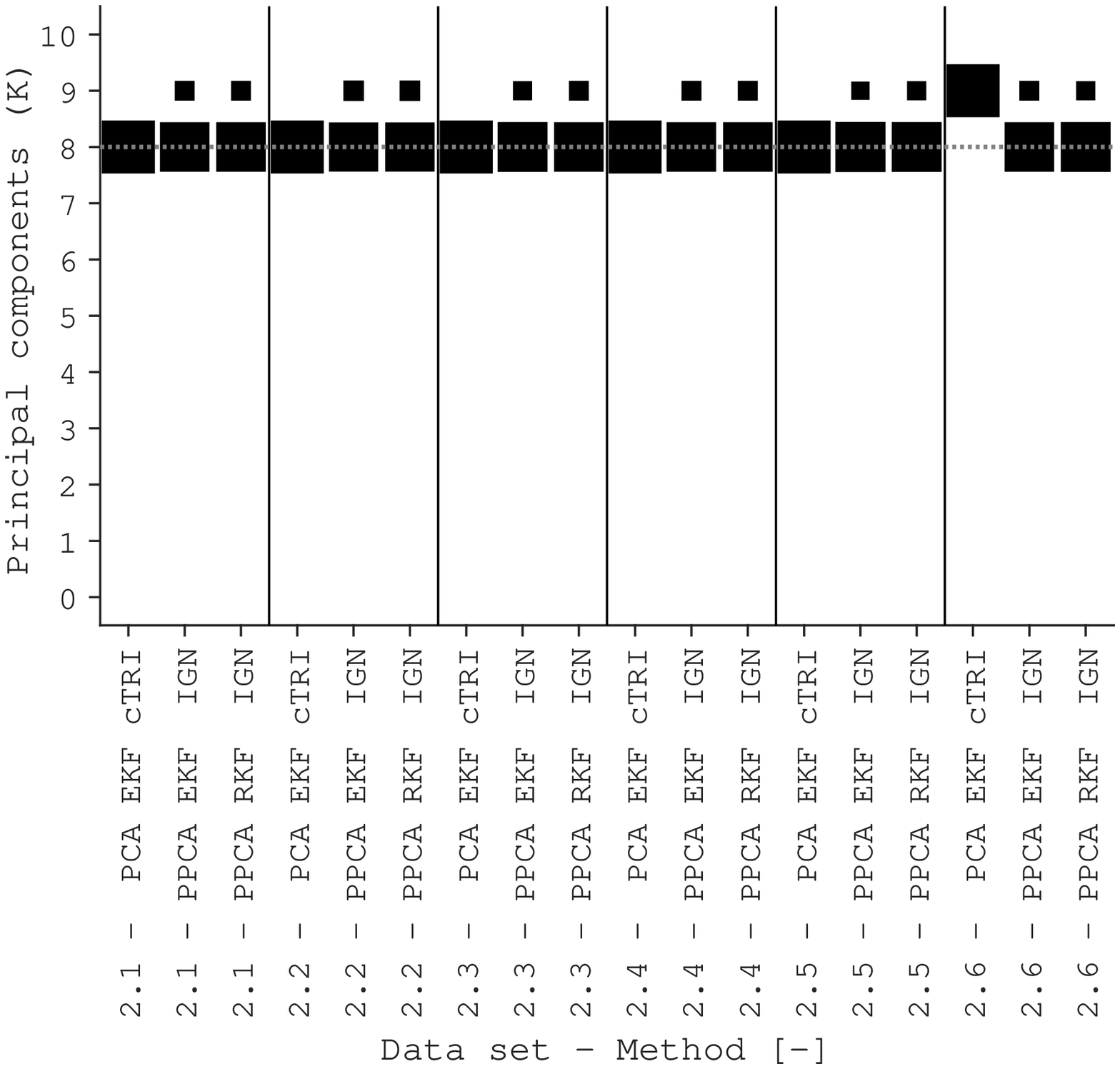}
		\caption{
{\bf Distribution of identified numbers of PCs for the simulated data sets 2.1 to 2.6.} For each number of principal components ( $K$, bottom to top), and for every noise level and every method (left to right), a black box is shown with a surface proportional to the number of data instances for which $K$ component are selected. The horizontal line indicates the simulated value of $K$ (8). With the PPCA-based methods, the identified number of PCs is never too low and at most 1 PC too high. With the PCA model, the results are perfect except for data set 2.6, where the number of retained PCs always equals the maximum value that could be obtained (9).
		}
		\label{fig:CVdist2}
	\end{figure}

\begin{figure}[H]
		\centering
		\includegraphics[width=1\textwidth, angle =+90]{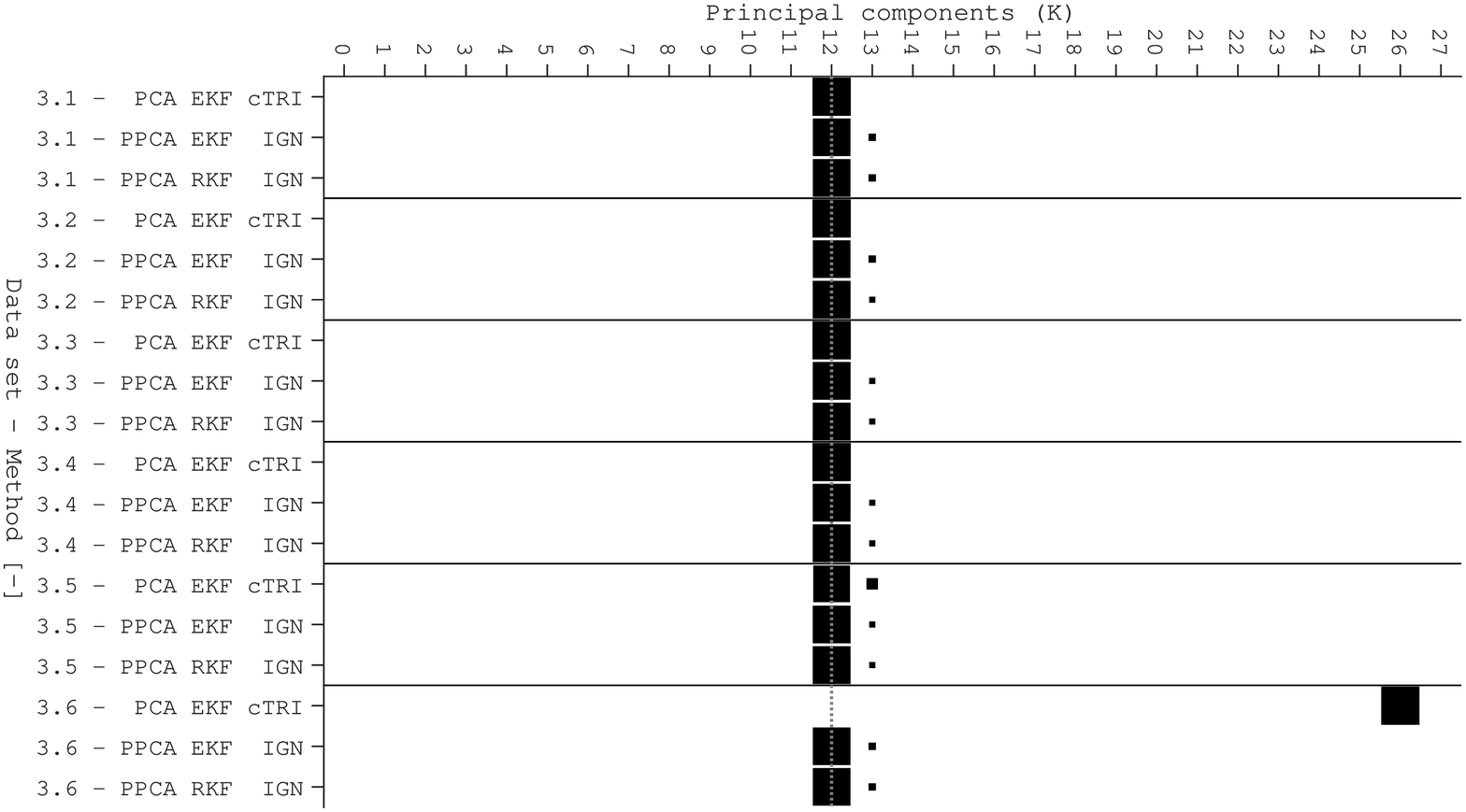}
		\caption{
{\bf Distribution of identified numbers of PCs for the simulated data sets 3.1 to 3.6.} For each number of principal components ( $K$, bottom to top), and for every noise level and every method (left to right), a black box is shown with a surface proportional to the number of data instances for which $K$ component are selected. The horizontal line indicates the simulated value of $K$ (12). With the PPCA-based methods, the identified number of PCs is never too low and at most 1 PC too high. With the PCA model, the results are perfect except for data set 3.5 and 3.6. With data set 3.6 and PCA EKF cTRI the number of retained PCs always equals the maximum value that could be obtained (9).
		}
		\label{fig:CVdist3}
	\end{figure}

\begin{figure}[H]
		\centering
		\includegraphics[width=1.1\textwidth, angle =+90]{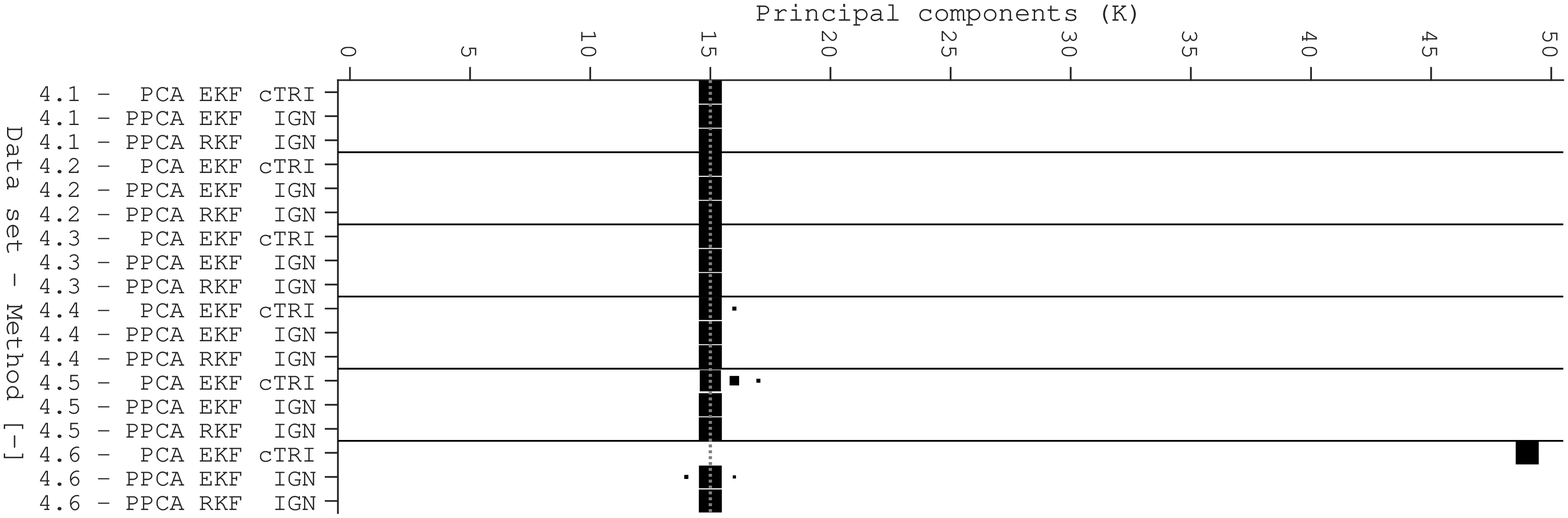}
		\caption{
{\bf Distribution of identified numbers of PCs for the simulated data sets 4.1 to 4.6.} For each number of principal components ( $K$, bottom to top), and for every noise level and every method (left to right), a black box is shown with a surface proportional to the number of data instances for which $K$ component are selected. The horizontal line indicates the simulated value of $K$ (1). With the PPCA-based methods, the identified number of PCs is at most 1 PC too low (set 4.6, PPCA EKF IGN) and at most 2 PCs too high. With the PCA model, the results are perfect except for data set 4.4 to 4.6. With data set 4.6 and the PCA EKF cTRI method, the number of retained principal component always equals the maximum value that could be obtained (49).
		}
		\label{fig:CVdist4}
	\end{figure}

\reffig{fig:CVdist1} shows the distribution of identified numbers of PCs for the data sets 1.1 to 1.6.  The PCA EKF cTRI method fails at noise level 6 by selecting the maximal number of PC that this method could deliver. In contrast, the PPCA-based methods also fail at reduced noise levels yet never identify a number of PCs that is more than 2 PCs too many. Similar results are obtained for data sets 2.1 to 2.6 (\reffig{fig:CVdist2}), 3.1 to 3.6 (\reffig{fig:CVdist3}), and 4.1 to 4.6 (\reffig{fig:CVdist4}). This suggests that the performance of the PPCA EKF IGN and PPCA RKF IGN methods degrades more gracefully than the PCA EKF cTRI method at high noise levels.

\reffig{fig:compute} displays the average computational effort, measured in seconds, required to complete a single run of each cross-validation method. Not surprisingly, the computational effort increases dramatically with the size of the data set. Little to no effects are visible for the noise level. Importantly, the PPCA RKF IGN method offers tangible computational benefits with computational requirements up to 100 times lower for the data sets with the largest number of variables (4.1 to 4.6). This also means that the PPCA RKF IGN method outperforms the PPCA EKF IGN method in both accuracy and computational efficiency for data set types 3 and 4, regardless of the applied measurement noise level. 

Note that the average run times remain low for all methods due to the use of a 16-fold cross-validation pattern in the row/sample direction. Initial experiments (not shown) suggest that the computational time depends linearly on the number of folds. A ball-park estimate for the run time when applying a leave-one-out pattern in the row direction can be obtained by multiplying the computed run times by 64. 

\addfig{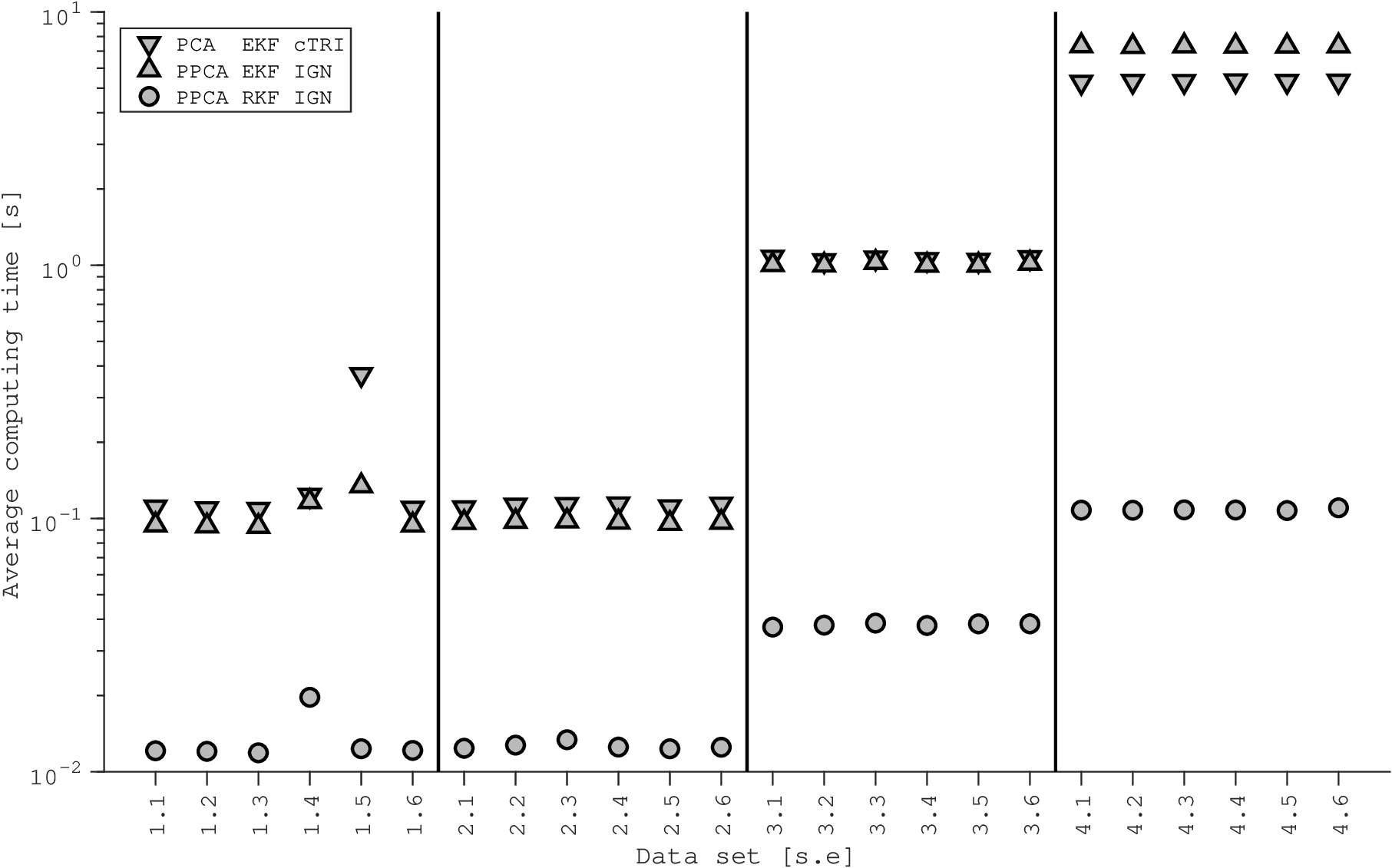}{{\bf Simulated data sets -- Computational requirements.} On average, cross-validation with PCA EKF cTRI and PPCA EKF IGN takes less than a second for data sets of type 1 and 2 while it takes under a minute for data sets of type 3 and 4. The PPCA RKF IGN is a factor 10 to 100 more efficient, with the average run times around a second for the largest data sets.}{fig:compute}

\subsection{Laboratory data set}

\subsubsection{Laboratory data set 0 -- Wavelengths 200-735 nm}

\reffig{fig:spectra} displays the complete set of absorbance measurements associated with the first validation block. One can see observe that the spectra are sensitive to variations of nitrite and nitrate concentrations in the range from 200 nm until about 420 nm. One can also clearly observe the known secondary absorbance peak of nitrate around 300 nm and of nitrite around 355 nm \cite{Spinelli2007}. On the left hand side of the spectra, i.e. below 250 nm, one can observe high but rather insensitive absorbance measurements. This is a region where the Beer-Lambert law for absorbance measurements does not apply as the device is subject to saturation phenomena, i.e. virtually all light in this wavelength range is absorbed leading to meaningless readings by the device \cite{Masic2015}. As a result, analyzing these data by PCA, let alone using these data to test the proposed cross-validation methods, makes little sense. For this reason, we continue below with an analysis based on a subset of the absorbance measurements which are expected to depend linearly on the nitrite and nitrate concentrations in the experimental solutions.

\addfig{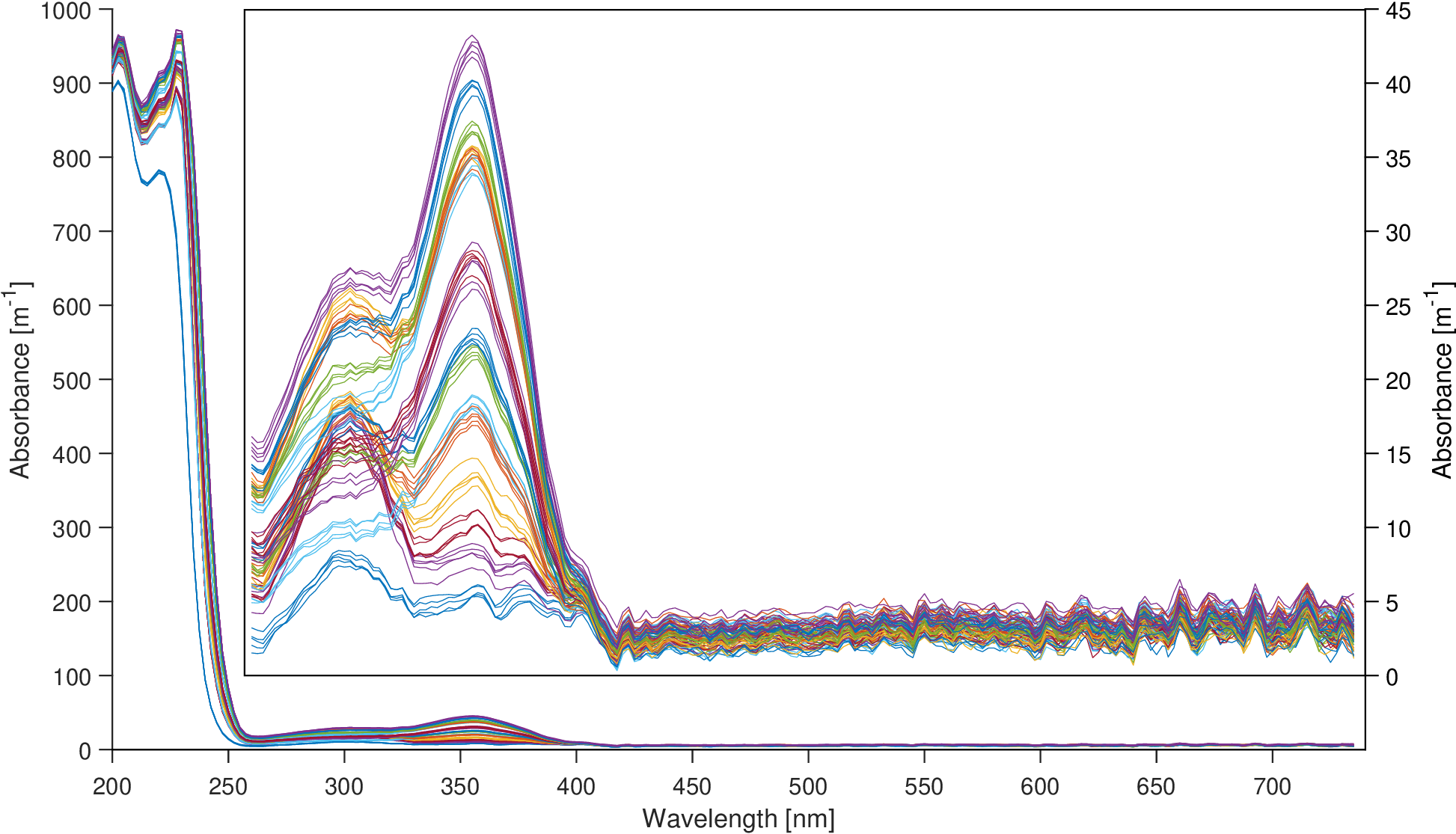}{{\bf Laboratory data set -- Raw data validation block 1.} Absorbance measurements are shown as a function of the wavelength. The inset on the top-right shows a detailed view on the measurements taken at 260 nm or higher wavelengths.}{fig:spectra}

\subsubsection{Laboratory data set 1 -- Wavelengths 285-735 nm}

The cross-validation methods are tested after the absorbance measurements corresponding to wavelengths below 285 nm are removed. \reffig{fig:uvvis1CV} shows the cross-validation results obtained with the resulting laboratory data set 1. With the PCA EKF cTRI method, the selected number of PCs is 21 whereas for the PPCA-based methods select 180 (PPCA EKF IGN) and 179 (PPCA RKF IGN) principal components. These estimated values are significantly higher than the anticipated number (2). 

\addfig{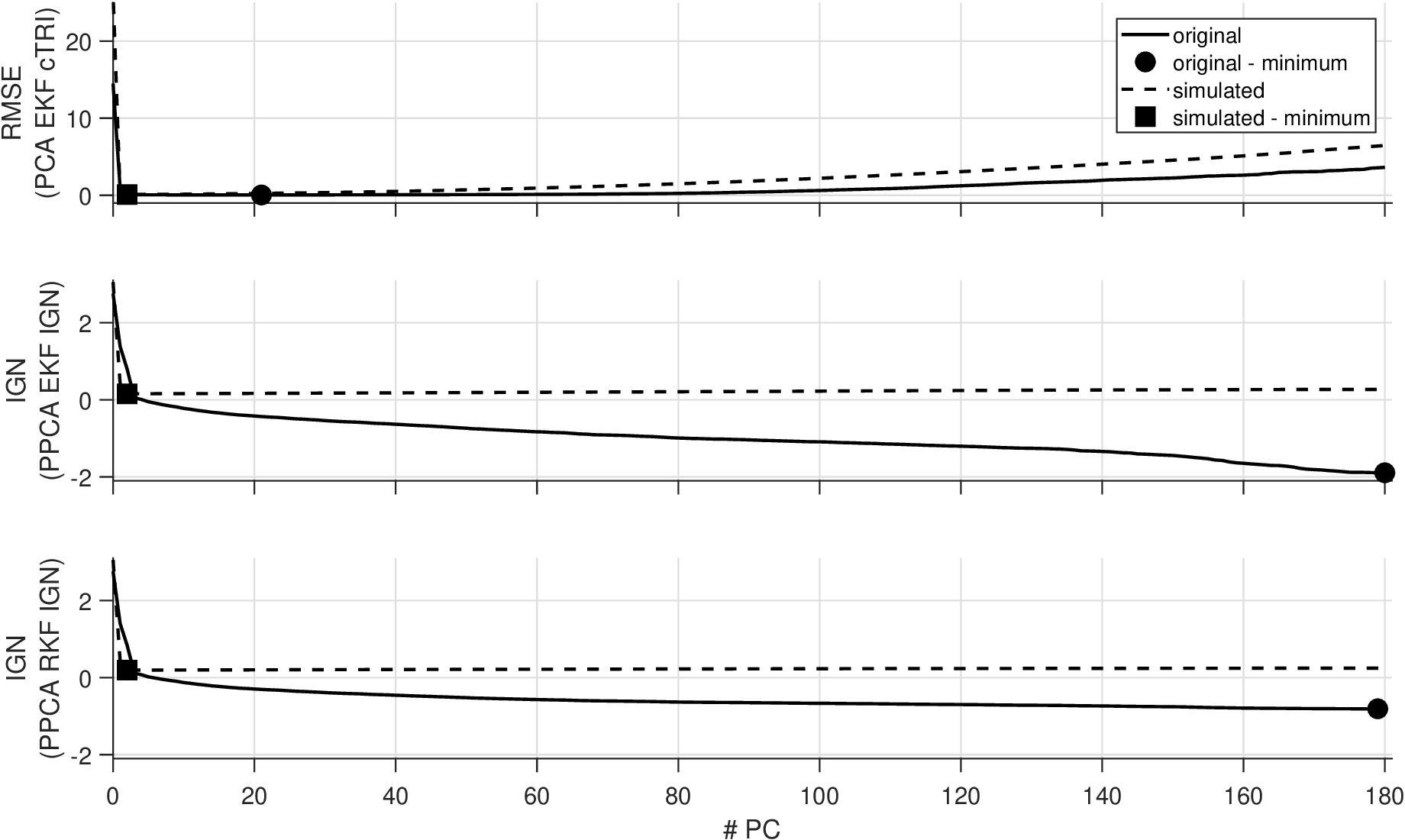}{{\bf Laboratory data set 1 -- Cross-validation results.} Cross-validated criteria as a function of the number of retained principal components. {\bf Top:} PCA EKF cTRI; {\bf Middle:} PPCA EKF IGN; {\bf Bottom:} PPCA RKF IGN; {\bf Full line:} Profiles with the original data; {\bf Dashed line:} Profiles with the simulated data matching PPCA model with 2 principal components; {\bf Circle:} Selected dimensionality with original data; {\bf Square:} Selected dimensionality with simulated data. With the original data, all cross-validation methods pick a number of principal components (circles) that is much higher (21, 180, 179) than expected on the basis of the experimental design (2). With simulated data, which is guaranteed to contain spherical noise, the selected number of principal components always matches the anticipated number (squares, 2).}{fig:uvvis1CV}

It is hypothesized that this result is caused by non-spherical noise present in the data, especially in the non-absorbing regions of the spectra (above 420 nm). To evaluate the merit of this explanation, 
\reffig{fig:uvvis10PC} shows the eigenvectors corresponding to the 3rd, 4th, and 5th PC. One can see that the 3rd PC appears to explain a uniform effect across the wavelengths in the visible range (400-735 nm). This cannot be explained by an effect of the absorbing nitrogen species and is therefore considered a strongly correlated type of noise. The scores for this PC were inspected visually to check for a temporal effect in the experimental data but none could be identified. The 4th and 5th PC appear to represent auto-correlated features with relatively large magnitudes in the red range of the spectrum (600-700 nm). This cannot be explained by an effect of nitrite or nitrate either and is therefore also considered an indicator of non-spherical type of noise. Similar patterns are observed for higher-order PCs as well ( $k\geq6$, not shown).

\addfig{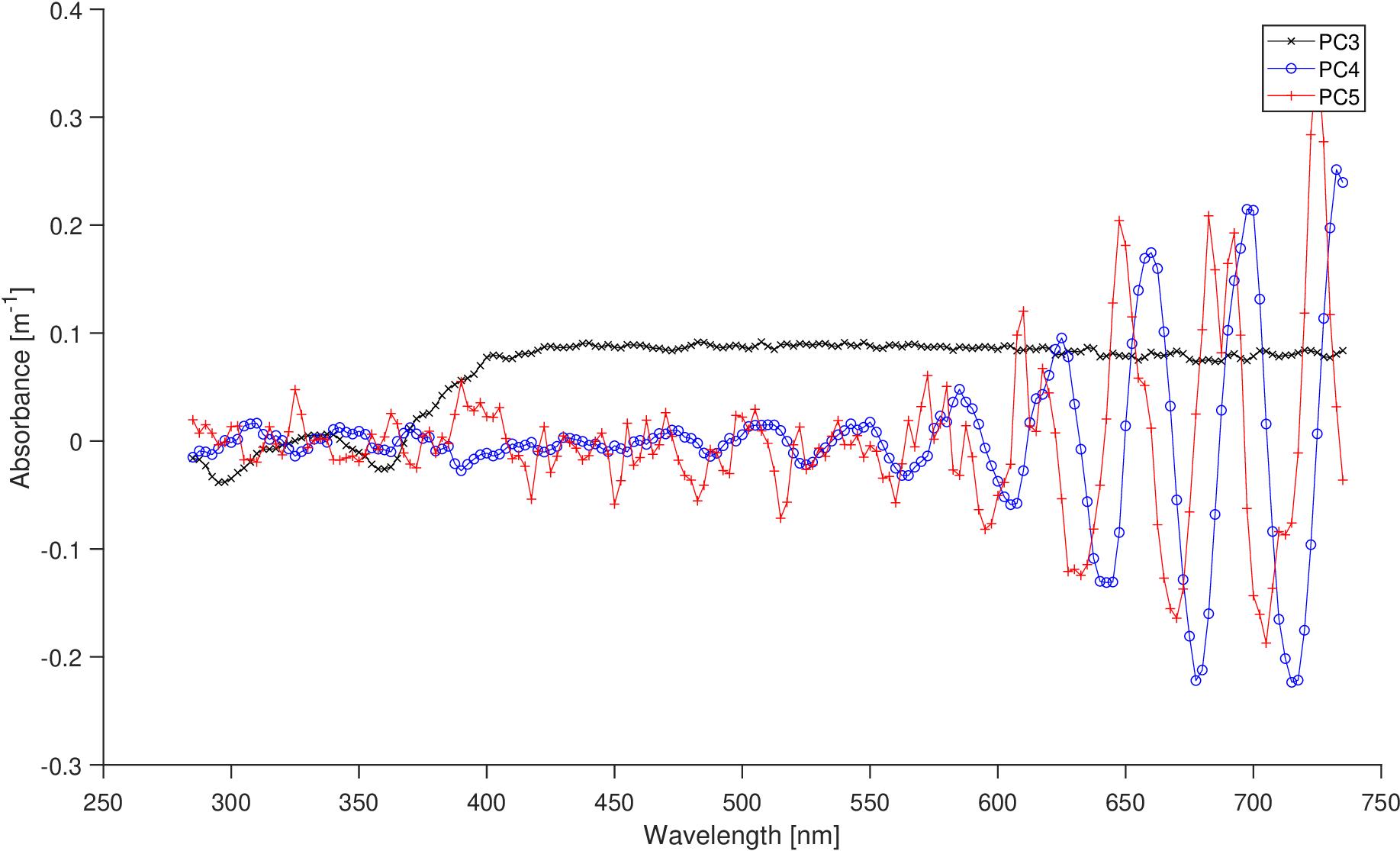}{{\bf Laboratory data set 1 -- Loading vectors for PC 3 to 5.} Cross-validated criteria as a function of the number of retained principal components. In this case, all cross-validation methods pick a higher number of principal components (circles) than expected (2).}{fig:uvvis10PC}

To corroborate that the identified non-spherical nature of the residuals is the leading cause for the failure to identify the correct number of PCs, we simulate data with a 2-PC PPCA model identified with the laboratory data set 1. That is, we identify the mean and $\Sigma^{(K)}$ with the complete data set and assuming $K=2$. We then simulate a new data set of the same dimensions according to \eqs{generation2}{generation3}. We use the computed scores obtained with \eq{deflatescores} to do this. Note that these scores do not adhere to a normal distribution by virtue of the square grid experimental design. The deflated principal scores for the simulated data thus closely approximate the same distribution for the scores while any non-spherical noise is avoided by design. We then repeat each of the proposed cross-validation methods on this artificial data set. The result of this is shown in \reffig{fig:uvvis1CV}. In this case, all methods select the anticipated number of PCs (2). Consequentially, it is plausible that the presence of non-spherical noise causes the studied cross-validation methods to fail. At the same time, this suggests that all of the methods are robust against deviations between the empirical distribution of the underlying  latent variables and the assumed multivariate normal distribution for these latent variables.

\subsubsection{Laboratory data set 2 -- Wavelengths 285-385 nm}

Considering that the non-spherical noise features identified above appear to be concentrated in the visible range of the absorbance spectra (400-750 nm), we now apply the selected cross-validation methods to the laboratory data set 2, which only contains absorbance measurements for wavelengths between 285 and 385 nm. \reffig{fig:uvvis2CV} shows the obtained results. In this case, the PCA EKF cTRI cross-method leads to the anticipated number of PCs (2). In contrast the PPCA-based methods select 40 PCs.  

\addfig{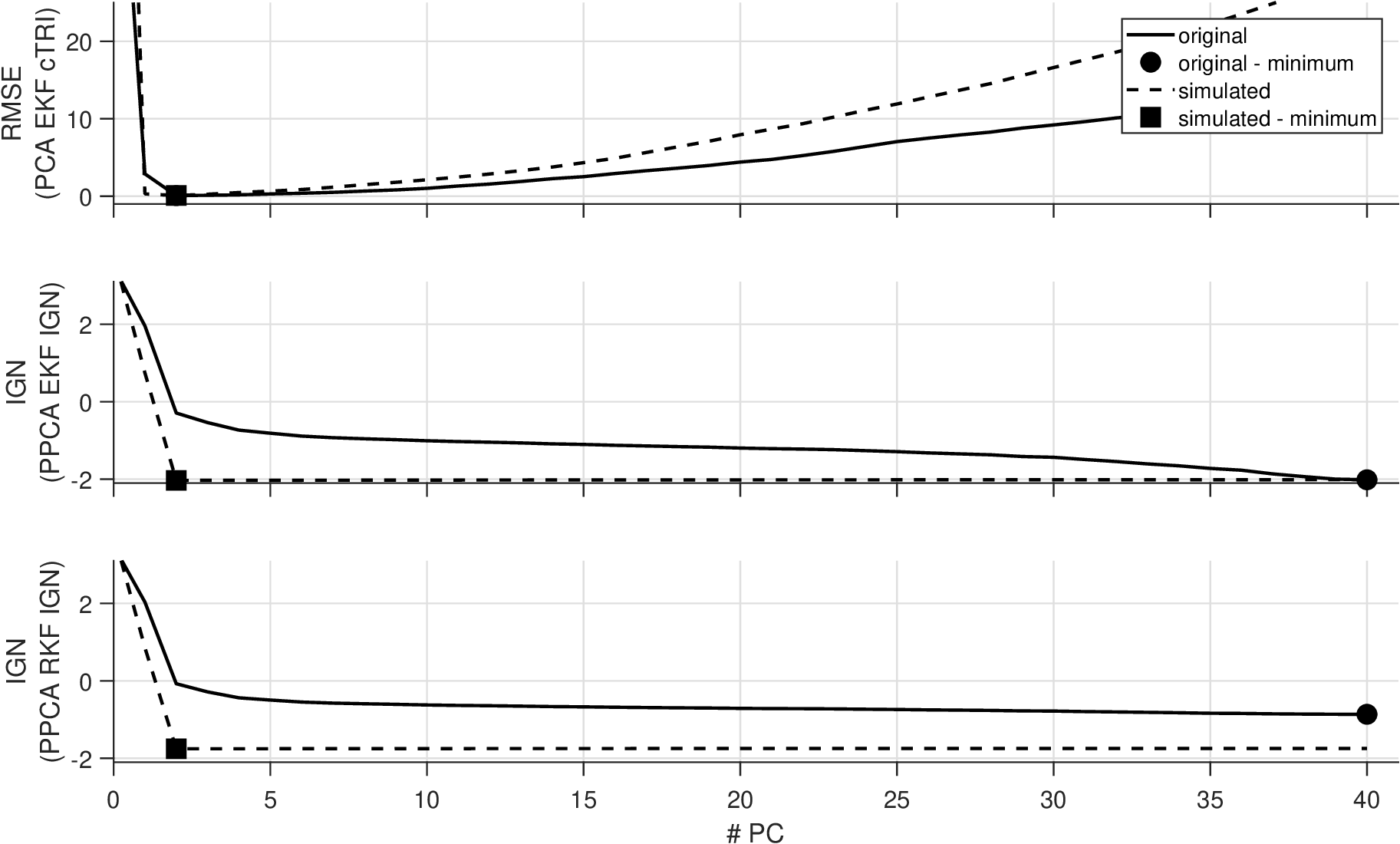}{{\bf Laboratory data set 2 -- Cross-validation results.} Cross-validated criteria as a function of the number of retained principal components. {\bf Top:} PCA EKF cTRI; {\bf Middle:} PPCA EKF IGN; {\bf Bottom:} PPCA RKF IGN; {\bf Full line:} Profiles with the original data; {\bf Dashed line:} Profiles with the simulated data matching PPCA model with 2 principal components; {\bf Circle:} Selected dimensionality with original data; {\bf Square:} Selected dimensionality with simulated data. With the original data, the PCA EKF cTRI methods picks the expected number of principal components, i.e. 2 (top panel, circle). In contrast, the PPCA-based methods pick 40 principal components (middle and bottom panel). With simulated data, which is guaranteed to contain spherical noise, every method selects the anticipated number of principal components (squares).}{fig:uvvis2CV}

A possible explanation is that the presence of non-spherical noise could not be removed entirely and that the PPCA-based cross-validation methods are more sensitive to this phenomenon. \reffig{fig:uvvis20PC} shows the eigenvectors associated with PC 3, 4, and 5. Each of these exhibits an oscillating profile, suggesting that auto-correlated noise is also present in the absorbance measurements in the ultraviolet range, i.e. where both nitrite and nitrate significantly contribute to the absorbance spectrum.

\addfig{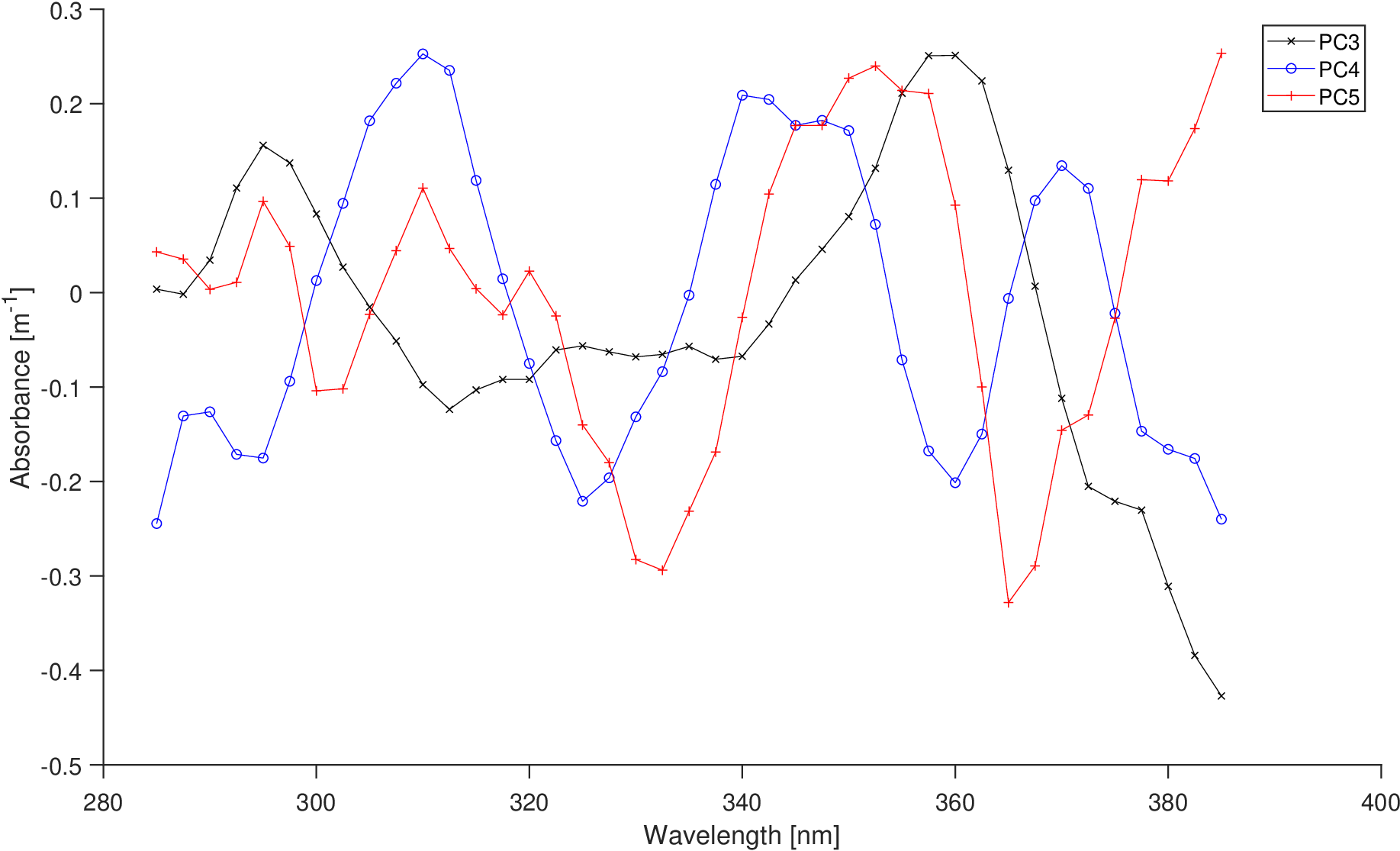}{{\bf Simulated laboratory data set 2 -- Loading vectors for PC 3 to 5.} Cross-validated criteria as a function of the number of retained principal components. In this case, all cross-validation methods pick a higher number of principal components (circles) than expected (2).}{fig:uvvis20PC}

As above, we apply the same cross-validation method with the laboratory data obtained through simulation of \eqs{generation2}{generation3} with a 2-PC PPCA model identified in the same way. The results are shown in \reffig{fig:uvvis2CV}. As before, the selected number of PCs equals the expected number, once more suggesting that {\em (i)} the presence of non-spherical noise is the main cause for a failure to identify the expected number of PCs with the PPCA-based methods and that {\em (ii)} the latent variables do not need to be distributed according to a multivariate normal distribution for successful model selection with the studied methods.

\section{Discussion}

\subsection{Cross-validated ignorance score as a tool for model identification}

With this study, we propose a new method for dimensionality selection in principal component analysis (PCA). It is based on the application of the ignorance score to the probabilistic principal component analysis (PPCA) model. Simulation results show that using the ignorance score delivers excellent accuracy in identifying the correct number of PCs when the assumptions of linearity and spherical noise are met. Most interesting is that this method clearly outperforms cross-validation based on element-wise cross-validation with corrected trimmed score imputation (PCA EKF cTRI) in the presence of high noise levels. In addition, the ignorance score can be applied successfully with a row-wise cross-validation (RKF) pattern, leading to the first known method which {\em (i)} delivers an accurate and meaningful minimum in the cross-validated performance criterion, {\em (ii)} is tuned well to the purpose of data compression, and {\em (iii)} has an efficiency which scales well with the dimensionality of the data set.

In the opinion of the authors, it is the absence of data leakage which explains this positive result. Indeed, unlike other methods for PCA model selection based on the RKF pattern, the proposed PPCA RKF IGN uses the validation data only during model testing and not during model prediction. 

\subsection{Cross-validated ignorance score as a tool for detection of model structure deficits}

Tests with a laboratory data set indicate that all considered methods select a number of principal components that is much higher than expected. Results are improved when variables that contain mostly noise are removed. A visual inspection of the identified PCA models and analysis with simulated data mimicking the laboratory data suggest that this can be explained by non-spherical  measurement error. Indeed, both PCA and PPCA are optimal only when the measurement errors are drawn from a spherical multivariate normal distribution. The presence of non-spherical measurement error produced by spectrophotometric devices for in-situ process monitoring has not been identified as a challenge before yet may explain why relatively complex models, i.e. with a large number of principal components, are necessary to obtain good predictive performance in practice \cite{Masic2015}. In the opinion of the authors, this means that the ignorance score is a viable tool to detect deviations from the assumed model structure, specifically the presence of non-spherical noise. 

\subsection{Link to existing work}

The ignorance score used for whole-sample validation was shown to be equivalent to the well-known Mahalanobis distance computed with the estimated variance-covariance matrix. This makes it similar -- yet not equal -- to the Mahalanobis distance based on the empirical variance-covariance matrix, as studied in \cite{Tong1995,DeMaesschalck2000,Brereton2015}, the Mahalanobis distance based on exploratory factor analysis \cite{Wu2006}, or the combined index composed of the Hotelling's $T^2$ statistic and the squared prediction error statistic \cite{Yue2001}. Logically, this means that ignorance score could be a useful statistic for anomaly and fault detection based on principal component analysis. It is also interesting to note that the combined index \cite{Yue2001} could be interpreted as a log-density, like the ignorance score, and is also sensitive to the selected number of principal components. Consequently, this combined index may also be useful for the purpose of dimensionality selection. Put together, this means that cross-validation with the ignorance score or the combined index can be used to identify the number of principal components that will minimize the expected value of such statistics, potentially leading to a predictable optimization of the trade-off between false alarm rates and true alarm rates in process monitoring applications.

\subsection{Open avenues for research}

In view of clarity, this work is focused on demonstrating the use of the ignorance score for dimensionality selection in the most trivial case for principal component analysis, i.e. assuming linear effects and spherical noise. However, the ignorance score is likely useful to determine the dimensionality of less restrictive models as well. For example, variational auto-encoders \cite{Dilokthanakul2016,Doersch2016} and Gaussian process latent variable models \cite{Lawrence2005} are interpreted as nonlinear versions of PPCA and permit a generative, probabilistic interpretation. Exploratory factor analysis (EFA) \cite{Joereskog1967} and target factor analysis (TFA) \cite{Bonvin1990,Harmon1995} may be used to find an optimal $K$-dimensional hyper-plane describing a data set similar to PCA, yet allowing for unequal noise variance estimates in the diagonal error variance-covariance matrix.

More recently, modified PCA models have been proposed to allow for non-diagonal forms for the error variance-covariance matrix. This matrix can be assumed known \cite{Hoefsloot2006}, estimated independently \cite{Wentzell1997,Wentzell1999}, or estimated simultaneously \cite{Narasimhan2015}. The eigenvalues associated with the residual space are however not set equal to each other in these models. This is unlike the EFA, TFA, and PPCA cases so that modifications would be necessary to obtain a probabilistic density model suited for evaluation and selection with the ignorance score. 

Other approaches to deal with non-spherical noise may consist of feature engineering prior to PCA analysis. For example, multi-scale principal component analysis \cite{Bakshi1998,Rosen2001,LeeDS2005} and functional principal component analysis \cite{Baert2012,Maere2012} are both based on the computation of new features, which typically are linear combinations of the original data prior to model calibration. This transformation may very well produce features that exhibit a noise variance-covariance matrix which is more diagonal than the original one. When so, this may improve the fit of the PPCA model and reduce the selected dimensionality. Conversely, specialized PCA models and feature generation should be explored as a way to enhance the robustness of the cross-validated ignorance score, specifically by accounting for nonlinear effects and for unknown or poorly understood noise properties. 

There are also several aspects of cross-validation with the ignorance score that remain interesting when the PPCA model structure is correct. For example, it is unclear what the effect is of the number of folds in the row-wise direction on the quality of model selection. In addition, it is unclear whether there is value in replacing cross-validation with bootstrapping to allow for resampling of validation samples.

\section{Conclusions}

This work addresses the important yet challenging selection of the optimal number of latent variables in principal component analysis (PCA). Two variations of a newly proposed method based on the cross-validated ignorance score are compared to an established and high-performing method based on the cross-validated imputation error with trimmed score imputation. Our findings are summarized as:
\begin{itemize}
    \item Simulation results show that our proposed method delivers a comparable performance with high-dimensional data sets with low noise levels and is preferable in presence of high noise levels. 
    \item Experimental results revealed that devices for in-situ measurement of spectrophotometric absorbance spectra in aquatic systems are prone to produce non-spherical measurement errors.
    \item The cross-validated ignorance score is a valuable addition to the tool set for both PCA model selection and detection of PCA model structure deficits.
\end{itemize} 

\begin{acknowledgement}
The authors would like to thank Karin Rottermann, Sylvia Richter, and Kai Udert for their contributions to the work presented in this paper. The study has been made possible by the Swiss National Foundation (project: 157097) and Eawag Discretionary Funds (grant number: 5221.00492.012.02, project: DF2018/ADASen).
\end{acknowledgement}

\begin{suppinfo}

The Supporting Information consists of a single package including:

\begin{itemize}
  \item Detailed procedures for data simulation
  \item Detailed procedures for experimental data collection
  \item Self-contained software which produces all results presented in this work
  \item All experimental data collected for the purpose of this study
\end{itemize}

\end{suppinfo}

\bibliography{PCA_ignorance_manu}

\end{document}